\documentstyle[11pt,times]{article}
\def\lsim{\mathrel{\rlap{\raise 2.5pt \hbox{$<$}}\lower 2.5pt}}
\def\gsim{\mathrel{\rlap{\raise 2.5pt \hbox{$>$}}\lower 2.5pt}}
\begin{document}
\bibliographystyle{plain}
\thispagestyle{empty}
\begin{small}
\begin{flushright}
IISc-CTS-2/97\\
BUTP-97/11\\ 
hep-ph/9707305\\
\end{flushright}
\end{small}
\vspace{-3mm}
\begin{center}
{\Large
{\bf Higher Threshold Parameters in \boldmath{$\pi\pi$} Scattering
}}
\vskip 0.5cm
B. Ananthanarayan\\
Centre for Theoretical Studies \\
Indian Institute of Science \\
Bangalore 560 012, India\\
\smallskip
and
\smallskip

P. B\"uttiker\\
Institut f\"ur Theoretische Physik\\
 Universit\"at Bern\\
CH--3012 Bern, Switzerland\\

\vskip 2cm

\end{center}
\begin{abstract}
A family of threshold parameters
which probe the stability of chiral predictions
is considered.   The relevant criteria
for the choice of threshold parameters
are discussed.   Sum rules for these quantities
are derived from dispersion relations and evaluated
from effective range formulae. 
Good agreement with two-loop chiral estimates
for many of these quantities is found and
interesting discrepancies are discussed.  
\end{abstract}

\medskip


\newpage

\noindent{\bf 1.} Dispersion relations for $\pi\pi$ scattering
amplitudes with two subtractions have been rigorously established
in axiomatic field theory \cite{gw1}.
It is
convenient to consider dispersion relations for
$s-$channel amplitudes of definite iso-spin, 
$T^I_s(s,t,u),\ I=0,1,2$, where $s,\ t,\ u$ are the
Mandelstam variables.
Furthermore, each of the amplitudes may be written down
in terms of a unique function as $T^0_s(s,t,u)=
3A(s,t,u)+A(t,u,s)+A(u,s,t)$, $T^1_s(s,t,u)=A(t,u,s)-A(u,s,t)$,
$T^2_s(s,t,u)=A(t,u,s)+A(u,s,t)$.
Unitarity, analyticity and crossing
symmetry have been used extensively to study this fundamental
process of elementary particle physics.  
Introducing a partial wave expansion for these amplitudes
via $T^I_s(s,t,u)=32 \pi \Sigma (2l+1) f_l^I(s) P_l((t-u)/(s-4))$,
elastic unitarity implies above threshold and below
the four-pion threshold that the partial wave amplitudes
may be described in terms of the phase shifts $\delta^I_l(s)$
by $f^I_l(s)=\sqrt{s/(s-4)} \exp (i\delta^I_l(s))\sin \delta^I_l(s)$,
where we have set the pion mass ($m_\pi=139.6$ MeV) to unity.  
Note the threshold
expansion ${\rm Re}f^I_l(\nu)=\nu^l(a^I_l+b^I_l \nu + c^I_l \nu^2+
d^I_l \nu^3+...), \ \nu=(s-4)/4>0$ being the square of
the three momentum in the centre of mass frame,
which defines the scattering lengths $a^I_l$, the
effective ranges $b^I_l$ and the higher threshold parameters
$c^I_l$ and $d^I_l$, etc.
Sum rules have been
established for (combinations of) scattering lengths
and effective ranges in the past employing analyticity and
crossing symmetry constraints.  One culmination of the
dispersion relation approach to $\pi\pi$ scattering has
been the Roy equations \cite{smr1,bgn} which is a system of coupled
integral equations for partial wave amplitudes which
further trades the two unknown t-dependent functions in
fixed-t dispersion relations for the scattering lengths
$a^0_0$ and $a^2_0$.  Roy equation fits to phase
shift data \cite{bfp,pp} have been
extensively studied. Best fits to data 
give $a^0_0=0.26\pm 0.05$ \cite{nagels}.

\bigskip

\noindent{\bf 2.} Chiral perturbation theory \cite{gl1} is the effective
low energy theory of the standard model and describes processes
involving pionic degrees of freedom, viewed as 
the approximate Goldstone
bosons of spontaneously broken 
axial-vector symmetry of massless QCD.
At leading order the $\pi\pi$ scattering amplitude
is given by the Weinberg result from PCAC: $A(s,t,u)=(s-1)/F_\pi^2$ \cite{sw1}.
(In chiral perturbation theory this amplitude has now been
computed to one-loop \cite{gl1} and
even two-loop order \cite{sternetal,bijnensetal}.)
Furthermore this implies that the only non-vanishing
threshold parameters are 
$a^0_0=7/(32 \pi F_\pi^2)$,
$a^2_0=-1/(16 \pi F_\pi^2)$,
$a^1_1=1/(24 \pi F_\pi^2)$,
$b^0_0=1/(4 \pi F_\pi^2)$ and
$b^2_0=-1/(8 \pi F_\pi^2)$.
It is important to note the well known result that
the set of functions: $t^0=(2\beta (s-4/3) + 5 \alpha/3)/(F_\pi^2),\
t^1=\beta(t-u)/(F_\pi^2),\ t^2=(-\beta(s-4/3)+2 \alpha/3)/( 
F_\pi^2)$, where
$\alpha$ and $\beta$ are arbitrary real constants, verifies dispersion
relations with two subtractions and vanishing absorptive
parts.  The Weinberg amplitude is a special case of
this general linear amplitude, with $\alpha=\beta=1$.
It may be noted that a generalized version of chiral
perturbation theory that is motivated by considerations
of a small quark condensate in the QCD vacuum allows
$\alpha$ to vary over a range between unity and as much
as three, and reorganizes the chiral power counting \cite{kms}.
In this discussion we confine ourselves to the more predictive
standard chiral perturbation theory with $\alpha=1$ for
much of our discussion.

Dispersion relations with two subtractions have been
used to write down sum rules for (combinations) of
some of these threshold parameters, for example the
Wanders sum rules which maybe written
down as:
\begin{small}
\begin{eqnarray}
18 a^1_1=2 a^0_0 - 5 a^2_0 
 -\frac{1}{4\pi^2}
\int_0^\infty \frac{d\nu}{\sqrt{(\nu(\nu+1))^3}}
\left[2\nu \sigma^0(\nu) - 3 (3\nu+2) \sigma^1(\nu) -5\nu\sigma^2(\nu)
\right],\nonumber \\
3 b^0_0=2 a^0_0 - 5 a^2_0 
 +\frac{1}{4\pi^2}
\int_0^\infty \frac{d\nu}{\sqrt{(\nu(\nu+1))^3}}
\left[(4\nu+3) \sigma^0(\nu) - 3 (2\nu+1) \sigma^0(0) -3 \nu \sigma^1(\nu)
+5\nu \sigma^2(\nu)\right],\nonumber \\
6 b^2_0=- 2 a^0_0 + 5 a^2_0 
 +\frac{1}{4\pi^2}
\int_0^\infty \frac{d\nu}{\sqrt{(\nu(\nu+1))^3}}
\left[2 \sigma^0(\nu) + 3 \nu \sigma^1(\nu) +(7\nu + 6)\sigma^2(\nu)
-6(2 \nu+1) \sigma^2(0)\right],\nonumber 
\end{eqnarray}
\end{small}
where $\sigma^I(\nu)\equiv (4\pi/\sqrt{\nu(\nu+1)})\cdot \sum (2l+1)
\mbox{Im }f^I_l(\nu)$ are the cross-sections.
The Weinberg predictions for the quantities involved in these
sum rules obey these relations identically with
all the cross-sections set to zero, since the original
amplitude obeys dispersion relations with vanishing absorptive
parts.  Furthermore, it may be noted that since these are
relations at leading order in the chiral expansion, it would
be fair to expect that the relevant dispersion integrals 
make numerically less significant contributions
to $18 a^1_1$, $3 b^0_0$ and $6 b^2_0$ compared to $2 a^0_0 - 5 a^2_0$,
even at higher orders in the chiral expansion. 
The presence of two subtractions in the dispersion relations
where $a^0_0$ and $a^2_0$ play effectively the role of
subtraction constants, render them difficult to pin down
on the basis of dispersion relation phenomenology alone:
{\it vital chiral inputs are required to make sharp predictions
for these quantities.}  

\bigskip

\noindent{\bf 3.} 
At one-loop order, chiral perturbation theory requires
the introduction of several coupling constants 
in $\cal{L}$$^{(4)}$,  which
account for the non-renormalizable character of the
non-linear sigma model Lagrangian, $\cal{L}$$^{(2)}$,
which is the basis of the Weinberg
result.  Four of these constants $\overline{l}_i,\ i=1,2,3,4$
enter $\pi\pi$ scattering and in particular $\overline{l}_1$
and $\overline{l}_2$ enter predictions for
$c^0_0, c^2_0, b^1_1, a^0_2$ and $a^2_2$,
while all higher threshold parameters will
receive no contributions from the trees generated
by $\cal{L}$$^{(4)}$
at this order.
In the past experimentally known values for 
$a^0_2$ and $a^2_2$ \cite{nagels} were used to fix these quantities
\cite{gl1}, while no values have been reported in the literature for
$c^I_l$ and $b^1_1$.  
For completeness we note the one-loop formulas
for $c^I_l$ since they have been reported earlier:
\begin{eqnarray}
c^0_0=\frac{1}{2304 \pi^3 F_\pi^4}(-295 + 88 \overline{l}_1+112
\overline{l}_2) \nonumber \\
c^2_0=\frac{1}{5760 \pi^3 F_\pi^4}(-193 + 40 \overline{l}_1+160
\overline{l}_2) \label{cI0}
\end{eqnarray}

Recently, rapidly converging
sum-rules were written down \cite{atw} in order to estimate $b^1_1$
which in principle could be used to fix the values
of the two chiral coupling constants of interest,
indeed, as one could from the values of $c^I_0$ should
they be known.   However, systematic ambiguities
inherent to one-loop predictions for these quantities
have also been discussed recently. $\overline{l}_1$
and $\overline{l}_2$ were fixed instead by rewriting
the chiral and axiomatic representations of the
scattering amplitudes to the appropriate order in
the momentum expansion \cite{ab}.  This method is now being
extended to the two-loop case in order to fix
the coupling constants that enter at that order
in CHPT.  Note that the trees generated by $\cal{L}$$^{(6)}$
will now contribute to the threshold parameters
$d^I_0, c^1_1, b^I_2$ and $a^1_3$.  
At the one-loop level they are given by the expressions:
$d^0_0=-1643/(40320 \pi^3 F_\pi^4),$
$d^2_0=-893/(40320 \pi^3 F_\pi^4),$
$c^1_1=-23/(13440 \pi^3 F_\pi^4),$
$b^0_2=-481/(201600 \pi^3 F_\pi^4),$
$b^2_2=-277/(201600 \pi^3 F_\pi^4)$ and
$a^1_3=11/(94080 \pi^3 F_\pi^4).$

To summarize,
we note that the trees generated by
$\cal{L}$$^{(2)}$, 
contribute to

\begin{eqnarray}
& \displaystyle a^I_0 \ \ \ b^I_0 & \nonumber\\
& \displaystyle a^1_1 & \label{treelevel}
\end{eqnarray}

\noindent while those generated by
$\cal{L}$$^{(4)}$ contribute to

\begin{eqnarray}
& {\displaystyle a^I_0 \ \ \ b^I_0 \ \ \ c^I_0} & \nonumber\\
& {\displaystyle a^1_1 \ \ \ b^1_1} & \label{oneloop} \\
& {\displaystyle a^I_2} & \nonumber
\end{eqnarray}

\noindent and those of 
$\cal{L}$$^{(6)}$
contribute to

\begin{eqnarray}
& \displaystyle a^I_0 \ \ \ b^I_0 \ \ \ c^I_0 \ \ \ d^I_0 & \nonumber\\
& \displaystyle a^1_1 \ \ \ b^1_1 \ \ \ c^1_1 & \label{twoloop} \\
& \displaystyle a^I_2 \ \ \ b^I_2 & \nonumber \\
& \displaystyle a^1_3. & \nonumber 
\end{eqnarray}

\bigskip

\noindent{\bf 4.}  The main purpose of this letter
is precisely to provide estimates to those quantities
appearing in eq.(\ref{oneloop}) and eq.(\ref{twoloop})
for which no information is available in the literature
and to compare whenever possible with the predictions
of CHPT.   Such a consistency check may be viewed as
a probe into the range of validity in energy of chiral
predictions.                      
Indeed, when Roy equation fits to the planned precision
experiments are performed, all the quantities discussed
here may be evaluated afresh, which would then amount to
a high precision experimental determination of these
numbers.  Such Roy equation fits may then be employed
to evaluate chiral parameters determined from dispersion
relation phenomenology, and such a consistency check may then
be performed again.

\bigskip

\noindent{\bf 5.}  We work in the approximation
that the absorptive parts are modeled entirely by
the S- and P- waves 
(In the numerical analysis we further assume that the
contribution to the dispersion integrals from the
S- and P- waves also above the $K\overline{K}$
threshold may be neglected as in Ref. \cite{atw}.)
which has been found to be justified phenomenologically
in the past and is supported today by chiral power
counting.
In this approximation it is particularly convenient to
represent the iso-spin amplitudes as \cite{bgn,mrw}:
\begin{eqnarray} 
& \displaystyle T^I(s,t,u)= 32 \pi \sum_{I'=0}^2 \left(
\frac{1}{4} (s {\bf I}^{II'} +t C_{st}^{II'}+
u C_{su}^{II'}) a_0^{I'} + \right. & \nonumber \\
& \displaystyle \left. \frac{1}{\pi} \int_{4}^{\infty}
\frac{dx}{x(x-4)} \left\{ \left[ \frac{s(s-4) {\bf I}^{II'}}
{x-s} +\frac{t(t-4)C_{st}^{II'}}{x-t}+\frac{u(u-4)C_{su}^{II'}}{
x-u} \right] {\rm Im} \ f_0^{I'}(x) \right. \right. & \label{amp} \\
& \left. \displaystyle \left. +3 \left[ \frac{s(t-u){\bf I}^{II'}}{x-s}+
\frac{t(s-u) C_{st}^{II'}}{x-t}+
\frac{u(t-s) C_{su}^{II'}}{x-u} \right] {\rm Im} \ f_1^{I'}(x) \right\}\right. .
& \nonumber 
\end{eqnarray}  
Projecting these amplitudes onto partial waves will
yield the Roy equations for each of the waves in the
S- and P- wave approximation.
[This approximation is equivalent to setting
the so-called {\it driving terms}
of the Roy equations to zero.  However, 
in the numerical work described below, we evaluate the contribution
of the $f_2(1270)$ resonance \cite{pdg} to the threshold parameters
of interest as a measure of the contribution of the
driving terms.]
The most convenient manner in which the sum rules of
interest may be computed is to consider the Roy equations
for each of the partial waves $f^I_0, \ f^1_1,\ f^I_2$
and $f^1_3$ in the neighborhood of the threshold, as
power series in $\nu$.
Furthermore,
in order to isolate the quantities of interest
we need to consider the Cauchy Principal Value of
the relevant integrals.  
The Principal Value singularity occurs due to self-coupling
of the waves and must be removed during the process of
computing the power series of the real parts of the waves.
An example is worked out for the $I=0$ S- wave in Appendix
A and the other waves may be treated analogously.  The
complete set of sum rules for the quantities of interest
that have not been published in the past is listed in Appendix B.

\bigskip

\noindent{\bf 6.} The spirit of this work will closely follow
that of Ref. \cite{mrw} wherein a modified effective range
formula was employed in order to model the absorptive
parts of the amplitudes.  
Note that the dispersion integrals we encounter
for (most of) the quantities of interest are very rapidly
converging which implies that it is the near
threshold region that needs to be modeled accurately,
the region where the effective range formula
is applicable.  
The modified effective range formula for the S- and
P- wave phase shifts
is of the type first proposed by Schenk \cite{schenk} and is:
\begin{eqnarray}
& \displaystyle \tan \delta^I_0(\nu)=\sqrt{{\nu\over \nu+1}}
\left\{ a^I_0 + [b^I_0-a^I_0/\nu^I_0 +(a^I_0)^3]\nu\right\} {\nu^I_0
\over \nu^I_0-\nu} & \nonumber \\
& \displaystyle \tan \delta^1_1(\nu)=\sqrt{{\nu^3\over \nu+1}}
\left\{a^1_1+[b^1_1-a^1_1/\nu^1_1]\nu\right\} {\nu^1_1
\over \nu^1_1-\nu}
\end{eqnarray}

\bigskip

\noindent{\bf 7.}  Our numerical work requires
inputs to the effective range formulae.  Much of
these inputs
are guided by one-loop chiral perturbation theory.
We also perform a sample computation with a
set of input parameters with $a^0_0$ corresponding to the
best fits to the experimental data of $K_{e4}$ of
0.26 and the rest of the quantities computed for that optimal
Roy equation fit. In Table 1 the complete set of inputs is
tabulated, besides the conventional choice $\nu^0_0=8.5,\,
\nu^1_1=6.6\, , \nu^2_0=-5.0$.  

In Table 2 we present the computed values of the threshold
parameters of interest for the inputs of Table 1.
The contribution of the $f_2$ resonance \cite{pdg} in the narrow
width approximation to the quantities of interest, calculated by
plugging in the appropriate absorptive part into the
relevant Roy equation and computed in the appropriate
limit, yields the results tabulated along $f_2$ in Table 2.  This should
be considered as setting the scale of the corrections
arising from all the higher waves and the higher energy
tail that would be described in terms of Regge parameterization
of the absorptive parts and possible Pomeron contributions.
It may be judged from this that the bulk of the contribution
is received from the S- and P- wave low energy absorptive parts.


In Table 3 we tabulate the one-loop and two-loop predictions
for these quantities whenever available \cite{sternetal,bm}.  
Let us first consider
the one loop predictions for $c^I_0$  which receive
contributions from $\overline{l}_1$ and $\overline{l}_2$.
The latter computed from optimal Roy equation fits of Ref. \cite{ab}
have been inserted into eq.(\ref{cI0}) to produce the
entries in the first two columns of Table 3.  The rest
of the quantities are of the pure one-loop variety and
their values numerically tabulated.  Alternatively, one
may take the values of the $c^I_0$ and solve for
$\overline{l}_{1,2}$ via. eq.(\ref{cI0}).  Taking
the extreme values [including the contribution
of the $f_2$] for $-0.0049\leq c^0_0\leq 0.010$ and $0.012\leq c^2_0\leq
0.015$ we obtain
$-4.7\leq\overline{l}_1\leq 0.08$ and $3.8\leq\overline{l}_2\leq 5.7$.
While these numbers are not
to be taken literally since the contributions of the
high energy tail are completely neglected and the
effective range formula is not a real substitute for
a Roy equation solution of the lowest partial waves,
they continue to provide an important consistency
check on the values of these coupling constants.
These are consistent with several prior determinations 
for these
quantities.  

The comparison of the values of $c^I_0$ with the
two-loop predictions continue to be encouraging
as we observe from Tables 2 and 3.  Some attention
may be paid to the quantity $c^1_1$:  at one-loop
order it is negative where as the two-loop prediction
is positive and larger in magnitude.  This value
creeps up towards the value provided by the sum rules.
It could be that the presence of the $\rho$ as
a non-perturbative feature of hadron dynamics is
responsible for the mismatch between even the two-loop
prediction and the sum rule result for $c^1_1$.


Furthermore, the Taylor series of ${\rm Re}f^I_l(\nu)=
\sqrt{{\nu+1 \over \nu}}
\sin(2\delta^I_l(\nu))/2$ computed
with the effective range formulae for the phase shifts, itself
yields coefficients $c^I_l$ and $d^I_0$ which are comparable
with the quantities computed from the sum-rules.  This
proves to be a check on the effective range parameterization
itself.  For instance, for the choice I in Table 1, we
find 
$c^0_0=0.0092,\ c^2_0=0.011,\ c^1_1=0.00092,\
d^0_0=-0.013,\ d^2_0=-0.0012$,  
whereas for that of
choice IV, we have
$c^0_0=-0.0025,\ c^2_0=0.011,\ c^1_1=0.00084,\
d^0_0=-0.015,\ d^2_0=-0.0013$.  

Thus we see that detailed considerations of unitarity, analyticity
and crossing in the near threshold region is in excellent agreement
with chiral predictions --- these, however, do not suffice to
discriminate between the standard and generalized scenarios of
chiral perturbation theory.

\bigskip

\noindent{\bf 8.}  In order to come to grips
with these numbers, we have also chosen to compare the
predictions for the P- wave threshold parameters with
numbers arising from resonance saturation with the $\rho$.
In particular, the narrow width formula for the
absorptive part generated by the $\rho$:
\begin{equation}
{\rm Im} f^1_1(x)=\pi \Gamma_\rho m_\rho \sqrt{{x\over
x-4}} \delta(x-m_\rho^2)
\end{equation}
gives rise to an amplitude $A(s,t,u)$ by inverting
eq.(\ref{amp}) via $A(s,t,u)=(T^0(s,t,u)-T^2(s,t,u))/3$,
\begin{equation}\label{rho}
A^\rho(s,t,u)={(48 \pi)\Gamma_\rho \over m_\rho^2 (m_\rho^2-4)^{3/2}}
({t(s-u) \over m_\rho^2-t}+{u(s-t)\over m_\rho^2-u}).
\end{equation}
This is identical to the result in eq. (C.9) of Ref. \cite{gl1}
obtained from effective Lagrangian techniques with the appropriate
identification of the relevant coupling constant when we take
into account the identities, $1/(m_\rho^2-t)=1/m_\rho^2\cdot(1+t/(m_\rho^2-t))$
and $[t(s-u)+u(s-t)]=[-2(s-2)^2+1/2(s^2+(t-u)^2)]$.  This is
a consequence of two subtractions in the dispersion relations
used here.  
The formula eq. (\ref{rho}) when inserted into the
$I=1$ amplitude and projected onto the P- wave yields for $a^1_1
-(2a^0_0-5a^2_0)/18=0.0073,\ b^1_1=0.0077,
\ c^1_1=0.0006$, when parameters of the resonance \cite{pdg}
are inserted into the formulas for these quantities.
Note the $\rho$
contribution to 
$\overline{l}_1$ and $\overline{l}_2$ are   
$-\pi m_\rho/(2\Gamma_\rho)$ and $\pi m_\rho/(4 \Gamma_\rho)$
respectively [see, e.g., eq. (C.10) in \cite{gl1}].

Another treatment of resonance saturation \cite{ppo} with the
$\rho$ 
is based on writing an unsubtracted dispersion relation for
the $I=1$ t-channel amplitude divided by $(s-u)$. This is
expected to converge under the assumption of the validity of
the Pomeranchuk theorem, supported by the
behavior of the relevant Regge trajectory.  
The $\rho$ contribution to $\overline{l}_1$ and $\overline{l}_2$
are, however, 
$-\pi m_\rho/(4 \Gamma_\rho)$ and $\pi m_\rho/(4 \Gamma_\rho)$
respectively
[as read off from eq. (28) and eq. (32) of \cite{ppo}].  
This is in variance with eq. (C.10) of \cite{gl1}.
This must result from considering dispersion relations without
a sufficient number of subtractions that guarantee the absence
of crossing constraints on the absorptive parts due to the P- wave.
[An analogous situation arises with the resonance saturation with states
of $l\geq 2$ of fixed-t dispersion relations with two subtractions; this
does not yield a crossing symmetric amplitude and must be
treated with care.]

\bigskip

\noindent{\bf 9.} In the preceding sections we have established
a program of comparison between families of threshold parameters
for which sum rules have been derived and the quantities numerically
estimated, and then compared with one- and two-loop predictions
for these quantities.   The role of the number of subtractions
in dispersion relations is crucial and has been emphasized in
a variety of ways.  Other recent efforts in this confrontation
and comparison between data, dispersion relations and chiral
perturbation theory use extrapolation to subthreshold regions
reaching analogous conclusions regarding the interplay of
analyticity and unitarity \cite{ppo2}.  Our work is also in keeping
with the expectation expressed in Ref. \cite{gkms} that sensitive
tests of QCD in low energy $\pi\pi$ scattering should use all
theoretical constraints and pertinent low energy observables.
It should, however, be emphasized that we have considered only
quantities that do not require vital chiral inputs such as
$a^I_0$:  work is in progress to extract sharp predictions for
these quantities.

After this work was completed, we have received a preprint \cite{gw2}
where several $O(p^6)$ coupling constants have been evaluated.
It would be important to study the implications of this evaluation
to the program discussed here and {\it vice versa}.

\bigskip

\noindent{\bf Acknowledgements:}  It is a pleasure to thank
H. Leutwyler for discussions and insights.  BA thanks B.~Moussallam and PB
thanks J.~Gasser for discussions.

\bigskip

\appendix
\section{Derivation of the Sum rules}
The three lowest partial waves can be written as: 
\begin{eqnarray}
   \mbox{Re }f^0_0(\nu ) & = & a^0_0 + (2 a^0_0 - 5 a^2_0)\frac{\nu}{3}%
                               \nonumber \\
                         &   & +4\sum^2_{I'=0}\sum^\infty_{l'=0}P\int^\infty%
                               _0 d\nu ' K^{l' I'}_{0 0}(\nu ,\nu ')
                               \mbox{Im }f^{I'}_{l'}(\nu ')\\
   \mbox{Re }f^1_1(\nu ) & = & (2 a^0_0 - 5 a^2_0)\frac{\nu}{18}%
                               \nonumber \\
                         &   & +4\sum^2_{I'=0}\sum^\infty_{l'=0}P\int^\infty%
                               _0 d\nu ' K^{l' I'}_{1 1}(\nu ,\nu ')
                               \mbox{Im }f^{I'}_{l'}(\nu ')\\
   \mbox{Re }f^2_0(\nu ) & = & a^2_0 - (2 a^0_0 - 5 a^2_0)\frac{\nu}{6}%
                               \nonumber \\
                         &   & +4\sum^2_{I'=0}\sum^\infty_{l'=0}P\int^\infty%
                               _0 d\nu ' K^{l' I'}_{0 2}(\nu ,\nu ')
                               \mbox{Im }f^{I'}_{l'}(\nu ')
%
\end{eqnarray}
As an example we give the derivation of the higher threshold parameters for
$I=0$, $l=0$.\\
Working in the $S$- and $P$-wave approximation, the summation over the angular
momentum $l$ runs from zero to one.  Besides, this is equivalent to
projecting eq.(\ref{amp}) on the relevant partial wave.
The contributions of the three lowest partial waves to $f^0_0(\nu)$ 
are explicitly written down as:
\begin{eqnarray}
   f^0(\nu) & \equiv & \frac{1}{\pi}P\int_0^\infty d\nu'
                        \left\{\frac{1}{\nu'-\nu}-\frac{2\nu}{3\nu'(1+\nu')}-%
                        \frac{3+5\nu'}{3\nu'(1+\nu')}+\frac{2}{3\nu}%
                        \ln\frac{\nu+\nu'+1}{\nu'+1}\right\}%
                        \mbox{Im }f^0_0(\nu'),\nonumber\\
   f^1(\nu) & \equiv & \frac{1}{\pi}P\int_0^\infty d\nu'
                        \left\{ -\frac{3(3\nu+2\nu'+4)}{\nu'(1+\nu')}+%
                        \frac{6(2\nu+\nu'+2)}{\nu\nu'}%
                        \ln\frac{\nu+\nu'+1}{\nu'+1}\right\}
                        \mbox{Im }f^1_1(\nu'),\\
   f^2(\nu) & \equiv & \frac{1}{\pi}P\int_0^\infty d\nu'
                        \left\{\frac{5(\nu-2\nu')}{3\nu'(1+\nu')}+%
                        \frac{10}{3\nu}\ln\frac{\nu+\nu'+1}{\nu'+1}\right\}
                        \mbox{Im }f^2_0(\nu'),\nonumber
\end{eqnarray}
such that
\begin{equation}
   \mbox{Re }f^0_0(\nu) = a^0_0 + (2 a^0_0 - 5 a^2_0)\frac{\nu}{3} +
                          f^0(\nu)+f^1(\nu)+
                          f^2(\nu).
\end{equation}
The singularity of the integrand resides in the first term of
$f^0(\nu)$. Adding and subtracting ${\rm Im} \, f^0_0(\nu')/(\pi(\nu+\nu'+1))$
does not change the integral and we may write
\begin{eqnarray}
  f^0(\nu) & = & G(\nu) + \nonumber\\
           &   &   \frac{1}{\pi}\int^\infty_0 d\nu'%
                        \left\{\frac{2}{3\nu}\ln\frac{\nu+\nu'+1}{\nu'+1}-%
                        \frac{2}{3\nu'(1+\nu')} - \frac{3+5\nu'}{3\nu'(1+\nu')}
                        -\frac{1}{\nu+\nu'+1}\right\}\mbox{Im }f^0_0(\nu'),
\end{eqnarray}

where
\begin{equation}
   G(\nu) = \frac{1}{\pi}P\int_0^\infty d\nu' \frac{2\nu'+1}{(\nu' 
-\nu)(\nu'+\nu+1)}%
                   \mbox{Im }f^0_0(\nu').\nonumber
\end{equation}
Consider the difference
\begin{eqnarray}\label{eq:nosing}
   f^0_0(\nu) - f^0_0(0) & = & (2 a^0_0 - 5 a^2_0)\frac{\nu}{3} +
                               G(\nu) - G(0) \nonumber\\
                         &    & + \bar{f}^0(\nu)-\bar{f}^0(0) +
                                f^1(\nu) - f^1(0) +
                                f^2(\nu) - f^2(0)\\
   \bar{f}^0(\nu) & \equiv & f^0(\nu) - G(\nu)\nonumber
\end{eqnarray}
We note that the
integrals not involving $G$ in eq.~(\ref{eq:nosing}) are free of singularities. 
Furthermore $G(\nu)-G(0)$ is also free of singularities as we show below.
Consider 
\begin{equation}
   G(\nu) - G(0) = \frac{1}{\pi}P\int_0^\infty\frac{(2\nu'+1)(\nu+\nu^2)}
                                 {(\nu'-\nu)(\nu+\nu'+1)(\nu'+1)\nu'}\mbox{Im 
}f^0_0(\nu')
\end{equation}
which still contains the Principal Value singularity. With
\begin{equation}
   P\int_0^\infty\frac{d\tau'}{\tau'^2 - \tau^2} = 0
\end{equation}
and $\tau' =\sqrt{\nu'(\nu'+1)}$ and $\tau =\sqrt{\nu(\nu+1)}$ we can write
\begin{equation}
   P\int_0^\infty d\nu'\frac{2\nu'+1}{\sqrt{\nu'(\nu'+1)}(\nu'-\nu)(\nu'+\nu+1)} 
= 0.
\end{equation}
Furthermore, in the S- (and P-) wave approximation, 
assuming normal threshold behavior, we have:
\begin{equation}
   \mbox{Im }f^0_0(\nu) = \frac{1}{4\pi}\sqrt{\nu(\nu+1)}\sigma^0(\nu)
\end{equation}
and therefore we can write
\begin{equation}
   G(\nu) - G(0) = \frac{1}{4\pi^2}\int^\infty_0 d\nu'\frac{(2\nu'+1)
                   (\nu+\nu^2)}{\sqrt{\nu'(\nu'+1)}(\nu'-\nu)(\nu+\nu'+1)}
                   \left\{\sigma^0(\nu')-\sigma^0(\nu)\right\}
\end{equation}
which is seen to be free from the Principal Value singularity.

The higher threshold parameters are now given by the
Taylor coefficients of the difference $f^0_0(\nu)-f^0_0(0)$.
\section{Sum rules for the higher threshold parameters}
For the higher threshold parameters considered in the text
and which have not been considered in the literature \cite{ab2},
 we find
\begin{eqnarray}
   c^0_0 & = & \frac{64}{\pi}\int_0^\infty d\nu\left[\frac{5}{288(1+\nu)^3}
               \mbox{Im }f^2_0(\nu) - \frac{1+2\nu}{32\nu(1+\nu)^3}
               \mbox{Im }f^1_1(\nu)\nonumber\right.\\
         &   & \left. + \left(\frac{1}{64\nu^3}+
               \frac{1}{288(1+\nu)^3}\right)\mbox{Im }f^0_0(\nu)%
               -\frac{\sqrt{\nu(\nu+1)}}{256\pi\nu^3(1+\nu)^3}(1+2\nu)\times
               \right.\\
         &   & \left. %
               \left\{\sigma^0(0)(1+\nu+\nu^2)+\frac{d}{d\nu}\sigma^0(\nu)|%
               _{\nu=0}\right\}
               \right]\nonumber\\[3mm]
   d^0_0 & = & \frac{256}{\pi}\int^\infty_0 d\nu\left[-\frac{5}{1536(1+\nu)^4}
               \mbox{Im }f^2_0(\nu)+\frac{2+5\nu}{512\nu(1+\nu)^4}\mbox{Im }
               f^1_1(\nu)\right.\nonumber\\
         &   & \left. + \left(\frac{1}{256\nu^4}-\frac{1}{1536(1+\nu)^4}
               \right)\mbox{Im }f^0_0(\nu) -\frac{\sqrt{\nu(1+\nu)}}
                {1024\pi\nu(\nu+1)}(1+2\nu)\right.\\
         &   & \left.     \left\{\frac{1+\nu+\nu^2}{\nu^2(1+\nu)^2}
               \frac{d}{d\nu}\sigma^0(\nu)|_{\nu=0}+
               \frac{1+2\nu+2\nu^2}{\nu^3(1+\nu)^3}\sigma^0(0)+
               \frac{1}{2(\nu+\nu^2)}\frac{d^2}{d\nu^2}\sigma^0(\nu)|_{\nu=0}
               \right\} \right]\nonumber
\end{eqnarray}
Similar one gets the threshold parameters for other waves:
\begin{eqnarray}
   c^1_1 & = & \frac{256}{\pi}\int_0^\infty d\nu \left[
               \frac{1}{2560(1+\nu)^4}\mbox{Im }f^0_0(\nu)
              -\frac{1}{1024(1+\nu)^4}\mbox{Im }f^2_0(\nu)\right.\\
        &    & \left. +\left(\frac{1}{256\pi}-\frac{2+11\nu}{5120(1+\nu)^4} 
               \right)\mbox{Im }f^1_1(\nu)
               -\frac{\sqrt{\nu(1+\nu)}}{3072\pi\nu(1+\nu)}
               \frac{1+2\nu}{\nu+\nu^2}
               \frac{d^2}{d\nu^2}\sigma^1(\nu)|_{\nu=0}\right]\nonumber\\[3mm]
   c^2_0 & = & \frac{64}{\pi}\int_0^\infty d\nu \left[
               \frac{1}{288(1+\nu)^3}\mbox{Im }f^0_0(\nu)+
               \frac{1+2\nu}{64\nu(1+\nu)^3}\mbox{Im }f^1_1(\nu)\right.
               \nonumber\\
        &    & \left.+\left(\frac{1}{64\nu^3}+\frac{1}{576(1+\nu)^3}\right)
               \mbox{Im }f^2_0(\nu)-\frac{\sqrt{\nu(\nu+1)}}
               {256\pi\nu^3(1+\nu)^3}(1+2\nu)\times\right.\\
        &    & \left.
               \left\{(1+\nu+\nu^2)\sigma^2(\nu)+(\nu+\nu^2)
               \frac{d}{d\nu}\sigma^2(\nu)|_{\nu=0}\right\}\right]
               \nonumber\\[3mm]
   d^2_0 & = & \frac{256}{\pi}\int_0^\infty d\nu\left[-\frac{1}{1536(1+\nu)^4}
               \mbox{Im }f^0_0(\nu)-\frac{2+5\nu}{1024\nu(1+\nu)^4}
               \mbox{Im }f^1_1(\nu)\right.\nonumber\\
         &   & \left.+ \left(\frac{1}{256\nu^4}-\frac{1}{3072(1+\nu)^4}\right)
               \mbox{Im }f^2_0(\nu) -
               \frac{\sqrt{\nu(1+\nu)}}{1024\pi\nu(1+\nu)}(1+2\nu)
               \times\right.\\
         &   & \left.\left\{\frac{1+\nu+\nu^2}{\nu^2(1+\nu)^2}
               \frac{d}{d\nu}\sigma^2(\nu)|_{\nu=0}+
               \frac{1+2\nu+2\nu^2}{\nu^2(1+\nu)^3}\sigma^2(0)+
               \frac{1}{2(\nu+\nu^2)}\frac{d^2}{d\nu^2}\sigma^2(\nu)|_{\nu=0}
               \right\}\right]\nonumber
\end{eqnarray}
In the same way one gets the threshold parameters for $I=0,l=2$, $I=2,l=2$: 
\begin{eqnarray}
   b^0_2 & = & \frac{1}{30\pi}\int^\infty_0 d\nu\frac{1}{\nu(1+\nu)^4}
               \left[-\nu\mbox{Im }f^0_0(\nu) + (3\nu -6)\mbox{Im }f^1_1(\nu)
               -5\nu\mbox{Im }f^2_0(\nu)\right]\\[3mm]
   b^2_2 & = & \frac{1}{60\pi}\int^\infty_0 d\nu \frac{1}{\nu(1+\nu)^4}\left[
               -2\nu\mbox{Im }f^0_0(\nu) - (3\nu - 6)\mbox{Im }f^1_1(\nu)
               -\nu\mbox{Im }f^2_0(\nu)\right]\\[3mm]
\end{eqnarray}

A check on the sum rules for $d^I_0,\, c^1_1$ and $b^I_2$
is to saturate the right hand sides with the lowest order
chiral phase shifts which should then yield the pure one-loop
formulae for these quantities due to the perturbative
unitarity of the chiral expansion.  The sum rules for
$c^I_0$ do not converge fast enough; one may consider
appropriate linear combinations with say the D- wave
scattering lengths $a^I_2$ in which the $\overline{l}_{1,2}$
dependence drop out and to saturate the corresponding
sum rules with the lowest order chiral phase shifts
and reproduce the one-loop formulae for these combinations.

\newpage

\begin{tabular}{| c|c|c|c|c|c|c|}\hline
\# & $a^0_0$ & $b^0_0$ & $a^2_0$ & $b^2_0$ & $a^1_1$ & $b^1_1$ \\ \hline
I & 0.19 & 0.237 & -0.04 & -0.074 & 0.035 & 0.006 \\ \hline
II & 0.20 & 0.236 & -0.036 & -0.074 & 0.035 & 0.006 \\ \hline
III & 0.21 & 0.236 & -0.035 & -0.075 & 0.035 & 0.006 \\ \hline \hline
IV & 0.26 & 0.231 & -0.021 & -0.076 & 0.036 & 0.006 \\ \hline
\end{tabular}
\begin{center}
{\bf Table 1.}  List of inputs for the effective range formula
\end{center}

\begin{small}
\begin{tabular}{|c|c|c|c|c|c|c|c|c|}\hline
\# & $c^0_0$ & $c^2_0$ & $d^0_0$ & $d^2_0$ & $c^1_1$ & $b^0_2$ & $b^2_2$ &
$a^1_3$ \\ \hline
I & 0.0084 & 0.012 & -0.016 & -0.0045 & 0.00080 & -0.00031 & -0.00028 &
0.000049 \\ \hline
II & 0.0061 & 0.013 & -0.016 & -0.0047 & 0.00084 & -0.00031 & -0.00029 &
0.000051 \\ \hline
III & 0.0041 & 0.013 & -0.016 & -0.0050 & 0.00087 & -0.00032 & -0.00030 &
0.000053 \\ \hline \hline
IV & -0.0049 & 0.015 & -0.017 & -0.0061 & 0.00096 & -0.00036 & -0.00035 &
0.000060 \\ \hline \hline
$f_2$ & 0.0016 & 0.00024 & -0.000030 & -0.0000304 & $5.7\cdot 10^{-6}$ &
0.000016 & $2.2 \cdot 10^{-6}$ & $3.2\cdot 10^{-7}$ \\ \hline
\end{tabular}
\end{small}
\begin{center}
{\bf Table 2.} Higher threshold parameters computed from sum rules
for the inputs of Table 1.  The contribution of $f_2$ is also listed.
\end{center}

\begin{tabular}{|c|c|c|c|c|c|c|c|c|}\hline
\# & $c^0_0$ & $c^2_0$ & $d^0_0$ & $d^2_0$ & $c^1_1$ & $b^0_2$ & $b^2_2$ &
$a^1_3$ \\ \hline
O-L & 0.0075 & 0.015 & -0.0066 & -0.0036 & -0.00028 & -0.00039 &
-0.00022 & 0.000020  \\
 & 0.0081 & 0.015 & & & & & &  \\
 & 0.0087 & 0.015 & & & & & &  \\ \hline
T-L(S) & 0.0068 & 0.013 & -0.017 & -0.0049 & 0.00032 & -0.00031 & -0.00031 &
0.000050 \\ \hline
T-L(G) & & & & & 0.00046 & -0.00034 & -0.00035 & 0.000058 \\ \hline
\end{tabular}
\begin{center}
{\bf Table 3.}  Table of values of quantities computed at one- and
two-loop (standard and generalized) order.  For the $C^I_l$
the quantities are computed from $\overline{l}_1$ and $\overline{l}_2$
listed in Table 1 of Ref. \cite{ab}.  The two-loop quantities are
from \cite{sternetal} and \cite{bm}.
\end{center}.


\bigskip


\begin{thebibliography}{abcdef}

\bibitem{gw1} For a comprehensive review, see, G. Wanders, Springer
Tracts in Modern Physics 57 (1971) 22, G. H\"ohler, ed.

\bibitem{smr1} S. M. Roy, Phys. Lett. B36 (1971) 353.

\bibitem{bgn} J-L. Basdevant, J. C. Le Guillou and H. Navelet,
Nuo. Cim. A7 (1972) 363.

\bibitem{bfp} J-L. Basdevant, C. D. Froggatt, Nucl. Phys. B72 (1974)
413.

\bibitem{pp} M. R. Pennington and S. D. Protopopescu,
Phys. Rev. D7 (1973) 1429.

\bibitem{nagels} M. M. Nagels et al., Nucl. Phys. B147 (1979) 189.

\bibitem{gl1} J. Gasser and H. Leutwyler, Ann. Phys. 158 (1984) 142.

\bibitem{sw1} S. Weinberg, Phys. Rev. Lett. 17 (1966) 616.

\bibitem{sternetal} M. Knecht et al., Nucl. Phys. B457 (1995) 513.

\bibitem{bijnensetal} J. Bijnens et al., Phys. Lett. B374 (1996) 210.

\bibitem{kms} M. Knecht and J. Stern,
Contribution to the 2nd DAFNE Handbook, p. 169;
M. Knecht, B. Moussallam and J. Stern,
Contribution to the 2nd DAFNE Handbook, p. 221. 

\bibitem{atw} B. Ananthanarayan, D. Toublan and G. Wanders,
Phys. Rev. D53 (1996) 2362.

\bibitem{ab} B. Ananthanarayan and P. B\"uttiker, Phys. Rev.
D54 (1996) 1125.

\bibitem{mrw} G. Mahoux, S. M. Roy and G. Wanders, Nucl. Phys.
70 (1974) 297.

\bibitem{pdg} Particle Data Group, M.R.~Barnett et al.,
Phys. Rev. D54, 1 (1996).

\bibitem{schenk} A. Schenk, Nucl. Phys. B363 (1991) 97.

\bibitem{bm} B. Moussallam, unpublished.

\bibitem{ppo} M. R. Pennington and J. Portol\`es, Phys. Lett. B 344
(1995) 399.

\bibitem{ppo2} M. R. Pennington and J. Portol\`es, Phys. Rev. D 55
(1997) 3082.

\bibitem{gkms} M. Girlanda, M. Knecht, B. Moussallam and
J. Stern, preprint, hep-ph/9703448

\bibitem{ab2} B. Ananthanarayan and P. B\"uttiker, Phys. Rev.
D54 (1996) 5501.

\bibitem{gw2} G. Wanders, hep-ph/9705323.

\end{thebibliography}
\end{document}